\documentclass{svjour2}                    
\smartqed  
\usepackage{graphicx}
\begin{document}

\title{Scaling of Entanglement Entropy for the Heisenberg Model on Clusters Joined by Point Contacts}


\subtitle{Possible Violation of the Area Law in Dimensions Greater than One}


\author{B. A. Friedman        \and
        G. C.  Levine 
}


\institute{B. A. Friedman \at
              Department of Physics \\
              Sam Houston State University\\
              Huntsville, Texas 77341-2267, USA\\
              Tel.: +936-294-1604\\
              Fax: +936-294-1585\\
              \email{phy\_baf@shsu.edu}           
           \and
           G. C. Levine \at
           Department of Physics and Astronomy\\
           Hofstra University\\
           Hempstead, New York 11549, USA\\
           Tel.:+516-463-5583\\
           \email{gregory.c.levine@hofstra.edu}    
}

\date{Received: date / Accepted: date}

\maketitle

\begin{abstract}
The scaling of entanglement entropy for the nearest neighbor  antiferromagnetic Heisenberg 
spin model is studied computationally for clusters joined by a single bond.  Bisecting the balanced three legged Bethe Cluster, gives a second Renyi entropy and the valence bond entropy which scales as the number of sites in the cluster.  For the analogous situation with square clusters, i.e. two  $L \times L$ clusters joined by a single bond, numerical results suggest that the second Renyi entropy and the valence bond entropy scales as $L$.  For both systems, the environment and the system are connected by the single bond and interaction is short range.  The entropy is not constant with system size as suggested by the area law.
\keywords{Entanglement entropy \and Area law\and Valence Bond 
Monte Carlo }
\end{abstract}

\section{Introduction}
This paper is a numerical investigation of entanglement entropy of the nearest neighbor isotropic Heisenberg  spin 1/2 model on clusters, in particular,  a Bethe cluster  and two $L \times L$ square clusters joined by a single bond.  We are studying the ground state quantum mechanical properties of these models and the coupling $J$ is taken to be the same on every bond (including the single bond joining the clusters), that is the interaction is antiferromagnetic.  Recall that for a large Bethe cluster and for the square lattice there is very good numerical evidence that there is long range antiferromagnetic order (suitably defined for a Bethe cluster), though to the best of our knowledge there is no proof.  The numerical methods used are spin wave theory \cite{song}, direct diagonalization and valence bond Monte Carlo \cite{sandvik,hastings}. There is no sign problem associated with the models so Monte Carlo is an effective numerical method.  The quantities calculated by valence bond Monte Carlo are the valence bond entropy \cite{alet,chhajlany} and the $n=2$ Renyi generalized entropy $S_2$\cite{hastings}.  Both these quantities are straightforward (given the techniques in ref. \cite{hastings} ) to calculate by valence bond Monte Carlo.  Note that $S_1$, the Von Neumann is not so easy to calculate, however, it is believed that the same essential physics is contained in $S_2$ (but see ref. \cite{chandran}).  Generically, we will refer to all these entropies as entanglement entropy.  We only use "balanced" clusters, where the number of even sites is equal to the number of odd sites, thus the ground state has spin 0 \cite{lieb} and the ground state can be represented by a superposition of valence bond states \cite{liang,sandvik}.

\begin{figure}
 \includegraphics[width=0.48\textwidth]{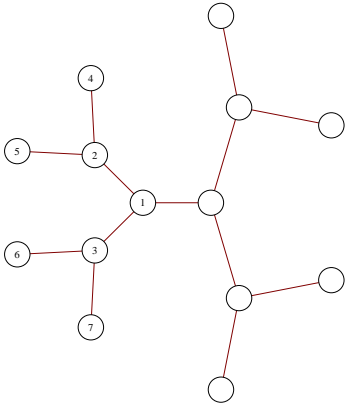}
\caption{14 site three branch Bethe cluster. }
\label{fig.1}
\end{figure}

In particular, for the present study one of the clusters we shall consider is the three branched bond centered Bethe clusters, see figure 1 for an illustration of the 14 site 3 branched bond centered cluster \cite{caravan}.  As discussed in \cite{caravan}, if you bisect such a cluster, a simple argument shows, the valence bond entropy must scale as the number of sites $N$.  In contrast, as we will later numerically demonstrate, the spin-wave technique gives an entropy proportional to $\log{N}$.  Because of  this apparent contradiction it is important  to use an unbiased numerical method, valence bond Monte Carlo, to calculate the Renyi generalized entropy $S_2$.  Due to recent advances in numerical techniques\cite{hastings,kallin} such calculations can be done accurately for large clusters.  A priori, however, one does not know if the clusters one can treat are large enough to see the asymptotic scaling law.  One of the objects of the current investigation is  to determine  if one can realize the asymptotic regime with existing numerical techniques and hardware.  

Why are we interested in the scaling law for entanglement entropy?  Firstly, if the entanglement entropy scales as $\ln{N}$, DMRG (Density Matrix Renormalization Group) is an effective, unbiased method to calculate the ground state properties. That is, one expects the number of states needed to describe the DMRG blocks to go as $e^{S_1} \approx O(N)$ not say $e^{N}$.  A still outstanding issue in condensed matter physics is how Neel order is destroyed by the addition of holes and ultimately is transformed into superconductivity.  Since the Bethe cluster has Neel order at 1/2 filling \cite{kumar,changlani} , the Hubbard or t-J model on the Bethe cluster would be an effective way to study this issue assuming the DMRG blocks scale with the number of sites, not the number of states,  in the blocks.  

Secondly, unlike in one dimension, the status of the area law for the entanglement entropy  is not as clear \cite{eisert}.  What are the conditions on the Hamiltonian and the cluster for the validity of the area law for a non one dimensional system?  Naively, since a single bond connects the two halves of the Bethe cluster, one would expect either a constant or a logarithmic dependence of entanglement entropy on system size.  However, for non interacting fermions, one sees $N$ dependent entropy for the Bethe cluster and $\sqrt{N}$ or $\log{N}$ (depending on boundary conditions) for square clusters separated by a single bond \cite{levine,caravan}.  Is this a pathological feature of non interacting systems?

There have been a number of very interesting papers \cite{bravyi,movassagh,aharonov,hastings2} where examples of models exhibiting large entanglement entropies or violating the area law, are investigated.  To best of our knowledge, the work developed in these papers does not apply directly to the particular systems we consider.  Broadly speaking, it seems in the above papers, the lattice or cluster is straightforward to experimentally realize, while the Hamiltonian is difficult to realize in a practical situation.  For our models, the Hamiltonian is physically realistic, however, a very large Bethe cluster is hard to realize in an experiment \cite{degennes}.  However, there is no such difficulty  in realizing the two $ L \times L$ cluster system.

\section{Bethe-Cluster}

\subsection{Spin-Wave Approach}

To calculate the Von Neumann entropy $S_1$ and the second generalized entropy $S_2$ from a spin-wave approximation we use the approach of \cite{song}  which is easily applied to the Bethe Cluster.  Let us consider initially a different situation from that considered in DMRG, namely, we take the subsystem inside the system.  As an example, consider in Figure 1, a three site subsystem, consisting of the sites labelled 1,2,3  inside the 14 site bond centered cluster.  Figure 2 is the Von Neumann entropy for subsystems of size 3 to 63 for the 254 and 510 site clusters.  

\begin{figure}
\includegraphics[width=0.48\textwidth]{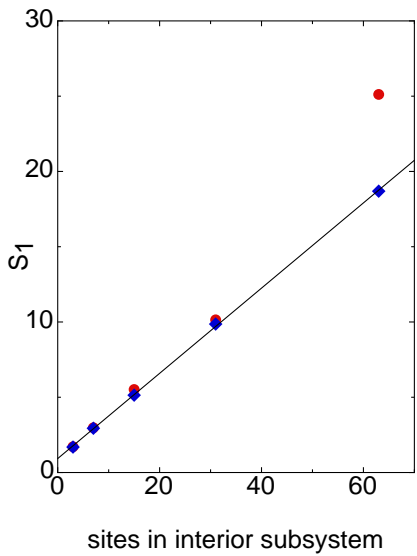}
\caption{$S_1$ vs. sites in the interior subsystem.  The red circles are for the 254 site cluster and the blue diamonds are for 510 sites.  The curve is a linear fit to the points for the 510 site cluster.}
\label{fig.2}
\end{figure}

We see the Von Neumann entropy scales with the number of sites in the interior cluster assuming one is sufficiently far  from the boundary of the cluster.  Note for the 254 site cluster the 63 site interior cluster is quite far from the linear fit in figure 2 while for the 510 site cluster the 63 site interior cluster is on the linear fit.  The linear scaling  is consistent with what one expect from the area law, since for an interior cluster, the number of boundary points scales with the number of sites in the cluster.  
   
Let us now examine a situation of greater similarity to that encountered in the blocking procedure in DMRG. Take a bond centered cluster (figure 1) and pick the subsystem to be the left half of the cluster.  

\begin{figure}
\includegraphics[width=0.48\textwidth]{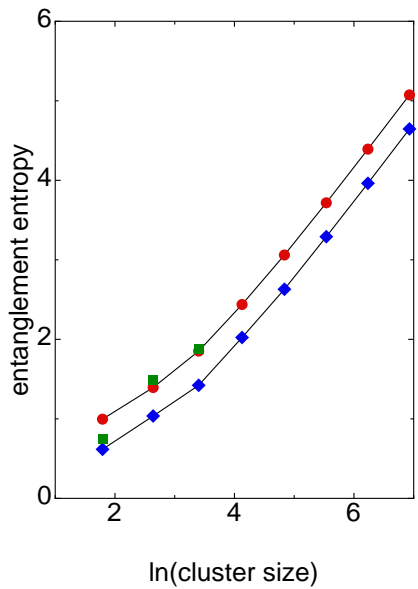}
\caption{Entanglement Entropy vs. ln(cluster size) for subsystems bisecting the system.  The red circles are for $S_1$ while the blue diamonds are for $S_2$.  The green squares are exact diagonalization results for  $S_2$.}
\label{fig.3}
\end{figure}

This is done in figure 3, the Von Neumann entropy and $S_2$ are plotted vs. the logarithm of the system size.  We see both quantities scale as the logarithm of the system size, similar to a one dimensional system. However, as previously mentioned, by the argument of ref. \cite{caravan}, the valence bond entropy scales as the number of sites.  Naively, one would expect, say the valence bond entropy and $S_1$ (or $S_2$) to scale the same way with system size ( at least up to logarithms \cite{kallin2}). Thus either the valence bond entropy scales differently from other entropies or the spin wave calculation gives an incorrect result.  In figure 3, the green squares refer to exact diagonalization results for $S_2$ for system sizes 6, 14, and 30.  The system sizes accessible to direct diagonalization (the 30 site cluster has state space of dimension approximately  $1.6 \times 10^8$ )  are too small to infer the scaling of $S_2$ with system size.  Nonetheless, some insight can be gained from plotting, in figure 4, the entanglement entropy for the 30 site cluster vs. the number of sites in the interior subsystem.  Perhaps not surprisingly, the entanglement entropy increases as the number of bonds connecting the subsystem to the system increases (up to 7 sites), after which the number of connections decreases and the entropy decreases.

\begin{figure}
\includegraphics[width=0.48\textwidth]{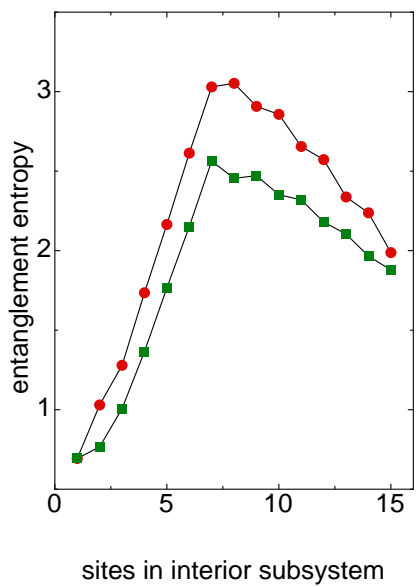}
\caption{Entanglement Entropy vs. number of sites in the interior subsystem for a 30 site cluster.  The red circles are for $S_1$ while the green squares are for $S_2$.  The points were calculated by exact diagonalization.}
\label{fig.4}
\end{figure}

\subsection{Valence Bond Monte Carlo}
We thus turn to valence bond Monte Carlo \cite{sandvik} as a method to compute, essentially exactly, the properties of the Heisenberg model on a Bethe Cluster.  In figure 5, the valence bond entropy is plotted vs. system size for a bisected system.  The valence bond entropy \cite{alet,chhajlany} is a natural quantity to compute with valence bond Monte Carlo, as the basis consists of valence bonds and the valence bond entropy counts the number of valence bonds leaving the subsystem.  From the figure, one sees that the valence bond entropy scales with system size.  Of course, given the argument in \cite{caravan}, this is no surprise.  Further insight can be obtained  by looking at the value of the valence bond entropy.  By the argument of \cite{caravan}, at least 1/3 of the valence bonds (for 1/2 the cluster, i.e. for a 1022 site cluster, roughly 170 bonds) must connect the two halves of the cluster; from figure 3, we see for large clusters, very close to 1/3 of the bonds connect the two halves.

\begin{figure}
\includegraphics[width=0.48\textwidth]{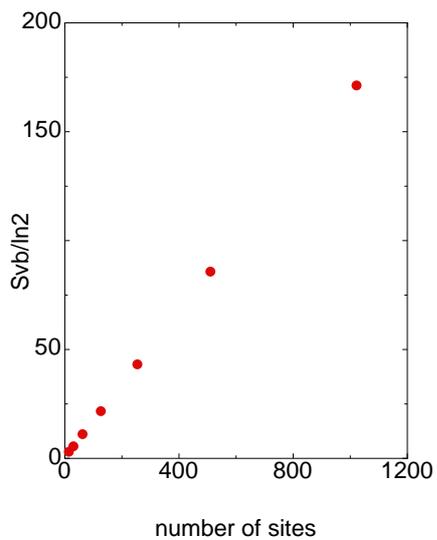}
\caption{Valence Bond Entropy vs. system size for a bisected system.}
\label{fig.5}
\end{figure}

Let us now consider the $n=2$ Renyi generalized entropy $S_2$.  Due to recent advances in computational technique this quantity can be calculated with valence bond Monte Carlo.  We apply the methods developed in ref. \cite{hastings}, see also \cite{kallin}; $S_2(\rho_A)$ is calculated as $-\ln{(\langle \rm{Swap}_A\rangle )}$ where A is the subsystem and $\rm{Swap}_A$ is a swap operator.  $\langle \rm{Swap}_A\rangle $ is calculated in the simplest formulation by a double projection Monte Carlo algorithm.  In a more sophisticated approach the ratios

\begin{equation}
\frac{\langle \rm{Swap}_{A^{i+r}}\rangle }{\langle \rm{Swap}_{A^{i}}\rangle }
\end{equation}

are computed from Monte Carlo; from these ratios $\langle \rm{Swap}_A\rangle $ is then calculated.  Here $r+i$ is a symbolic notation for a subsystem bigger than the subsystem $i$. For system sizes 6,14, 30 and 62 sites we have used three "different" approaches: the naive (no ratio) approach, the sophisticated approach where $i+r$ consists of the next shell in the Bethe cluster and the brute force "sophisticated" approach where "r" is only one site.  Recall there is a shell or layer  structure for Bethe  clusters, i.e. for the 30 site cluster (take 1/2 the cluster) the first layer  has 1 site, layer 2 has 2 sites, layer 3 has 4 sites, layer 4 has 8 sites. All three approaches agree, to within the statistical errors.  Exact diagonalization results  for 6,14 and 30 site clusters are also in agreement with the Monte Carlo calculations.

\begin{figure}
\includegraphics[width=0.48\textwidth]{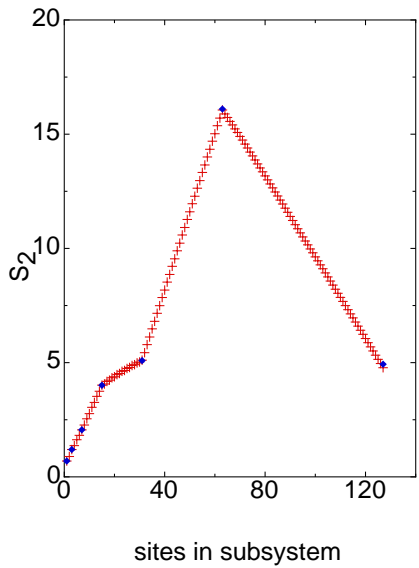}
\caption{$S_2$ vs. sites in the subsystem for a 254 site cluster.  The red crosses are calculated from the $r=1$ ratio method, while the blue diamonds use the shell technique.  Statistical errors are smaller than the symbols in the figure.   }
\label{fig.6}
\end{figure}

In figure 6, we plot $S_2$ vs sites in the subsystem for a 254 site cluster.   The subsystem is taken to be an interior subsystem as in figure 4.  The red crosses are calculated from the r=1 ratio method, while the blue diamonds use the shell technique.  Statistical errors are smaller than the symbols in the figure.  We see that $S_2$ appears to be a continuous piece wise linear function of the number of sites in the subsystem.  It is linear within the shell with a kink in going from one shell to another.  The kink (discontinuity in the derivative) is smaller in the shells near the center of the cluster.  The decrease in entropy for the final shell is presumably caused by the spins on the boundary of the cluster that are no longer connected by bonds  to spins outside the subsystem. 

\begin{figure}
\includegraphics[width=0.48\textwidth]{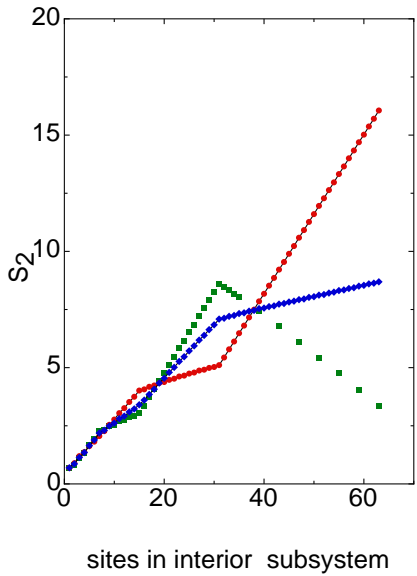}
\caption{$S_2$ vs. sites in the subsystem.    Green squares are for 126 site system, red circles are for  254 site system and blue diamonds are for 510 site system.  Statistical errors are smaller than the symbols in the figure.   }
\label{fig.7}
\end{figure}

We next consider a situation similar to figure 2.  In figure 7, we take an interior subsystem and calculate $S_2$ via valence bond Monte Carlo for systems of size 126 sites , green squares, 254 sites, red circles and 510 sites, blue diamonds.  Error bars would be smaller than the symbols in the figure.  It appears that $S_2$ scales with number of sites in the interior subsystem for a sufficiently small subsystem relative to the system as expected from the area law.

\begin{figure}
\includegraphics[width=0.48\textwidth]{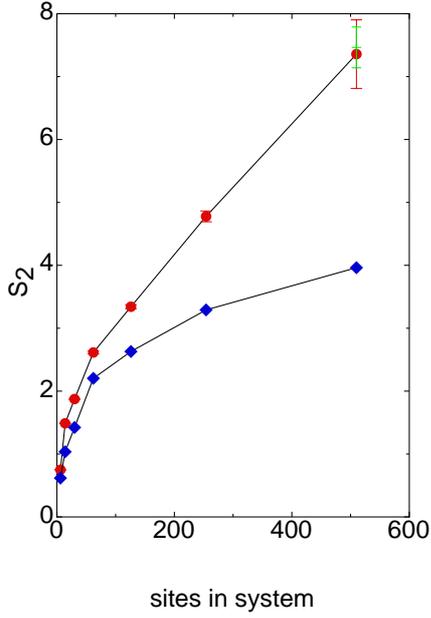}
\caption{$S_2$ vs. sites in the system for subsystems bisecting the system.  The blue diamonds are the spin wave results, while the red circles are calculated by valence bond Monte Carlo, $r=1$.  The green cross is for the shell method but breaking up the large shells to 32 sites.}
\label{fig.8}
\end{figure}

Finally in figure 8, we plot $S_2$ vs. system sizes for subsystems that consist of half the system.  The blue diamonds are the spin wave results, while the red circles are calculated by valence bond Monte Carlo, r=1.  The green cross is for the shell valence Monte Carlo method but breaking up the large shells to 32 sites.  For system sizes less than or equal to 62, at least qualitatively, Monte Carlo and spin wave agrees; however as the system sizes grow past this point there is an increasing discrepancy.  It appears the spin wave result increases logarithmically with the number of sites while the Monte Carlo results give linear dependence on the number of sites in the system.  Does this crossover with system size make sense?

One can present a simple minded argument that rationalizes this behavior.  The relevant "aspect" ratio $\alpha$ is the ratio of the number of generations to the number of boundary points, for example for the 6 site cluster $\alpha =\frac{2}{2} $, for 14 sites $\alpha=\frac{3}{4}$  for 30 sites $\alpha=\frac{4}{8}$ etc. .  One would expect $\alpha$ must be small to see Bethe cluster rather than 1 dimensional behavior.  Hence one would anticipate one needs to study system sizes substantially larger than 30 sites.

A heuristic argument can be made for the magnitude of the slope of $S_2$ vs. system size.  Recall that $S_2 = -\ln \langle \phi | \rm{Swap}_A|\phi\rangle $ where $\phi=|\psi_0\rangle  |\psi_0\rangle $ i.e. the replicated ground state.  Note that $|\psi_0\rangle =\sum_{\alpha} f_{\alpha} |\alpha\rangle $ where $|\alpha\rangle $ is a valence bond state and $f_{\alpha} \ge 0$.  Then the normalization 
\begin{equation}
1= \langle \phi | \phi\rangle  = \sum_{\alpha,\beta,\alpha',\beta'} f_{\alpha}f_{\beta}f_{\alpha'}f_{\beta'} \langle \beta'|\langle \alpha'|\alpha\rangle |\beta\rangle 
\end{equation}
and
\begin{eqnarray}
\langle \phi|\rm{Swap}_{A}|\phi\rangle  =
\\ \sum_{\alpha,\beta,\alpha',\beta'} f_{\alpha}f_{\beta}f_{\alpha'}f_{\beta'} \langle \beta'|\langle \alpha'|\rm{Swap}_{A}|\alpha\rangle |\beta\rangle 
\end{eqnarray}

All the valence bond states have at least $N/6$ bonds connecting the two halves of the Bethe cluster.  Typically, $\frac{N}{6}\frac{1}{3}\frac{1}{3}$ sites in A and an equal number in B are connected by valence  bonds in both $|\alpha\rangle $ and $|\beta\rangle $. 
Assume states $\langle \beta'|$  and $\langle \alpha'|$ with the same valence bonds dominate the normalization and the expectation value of $\rm{Swap}_{A}$.  The term $\langle \beta'|\langle \alpha'|\rm{Swap}_{A}|\alpha\rangle |\beta\rangle $ is reduced in comparison to $\langle \beta'|\langle \alpha'|\alpha\rangle |\beta\rangle $ by a factor of  $2^{-\frac{N}{6}\frac{1}{3}\frac{1}{3}}$ due to these bonds.  Hence
\begin{equation}
S_{2}\approx -\ln 2^{-\frac{N}{54}}=\frac{N}{54} \ln 2
\end{equation}

i.e.  $S_{2}$ is proportional to $N$ with a slope of about 1/78.  This is comparison  to the numerical results in figure 8 which give a slope of roughly $\frac{1}{100}$.

\section{Two Square Clusters Joined by a Single Bond}

\begin{figure}
\includegraphics[width=0.24\textwidth]{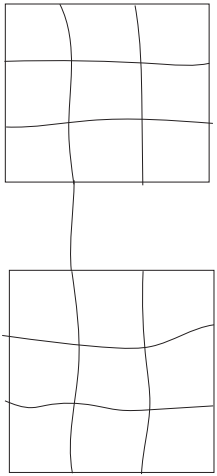}
\caption{4X4-4X4 cluster}
\label{fig.9}
\end{figure}

As in \cite{caravan} , consider two $L \times L$ clusters linked by a single bond. Here the bond is a spin-spin interaction of strength $J$, the same strength as all the other spin-spin interactions.  The single bond is chosen to be as close to the middle of the side as possible.  For $L$ even (odd) this bond joins the $L/2$ ( $(L+1)/2$) site on the side facing the other cluster with the corresponding site on the other cluster (figure 9).  There are several interesting aspects of this model.  Firstly, this model can be realized in a physical system, either solid state or cold atom.  Secondly, for $L$ even, there exists valence bond patterns where no valence bond need to go from one $L \times L$ cluster to the other.  Lastly, for non interacting fermions, at 1/2 filling, the entanglement entropy scales as $L$, not as a constant or $\ln L$.  This is consistent with the Bethe cluster results, both for free fermions and the Heisenberg model, in that the "perimeter" of the model (the sites with only one bond) scales as the number of sites in the cluster. 

\begin{figure}
\includegraphics[width=0.48\textwidth]{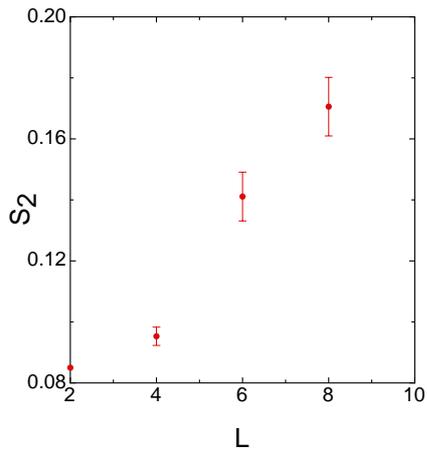}
\caption{$S_2$ vs $L$ for $L \times L$ - $L \times L$ clusters, $L$ even}
\label{fig.10}
\end{figure}

\begin{figure}
\includegraphics[width=0.48\textwidth]{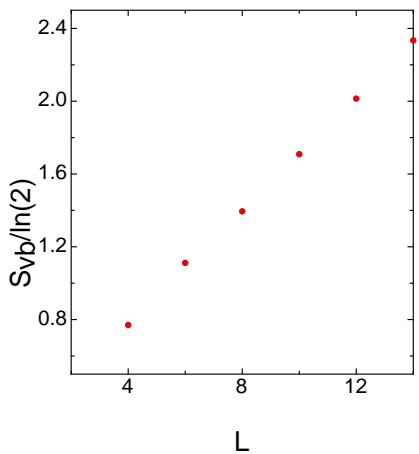}
\caption{Valence Bond Entropy vs. system size for a bisected system, $L$ even. Error bars are smaller than the data symbols.}
\label{fig.11a}
\end{figure}

\begin{figure}
\includegraphics[width=0.48\textwidth]{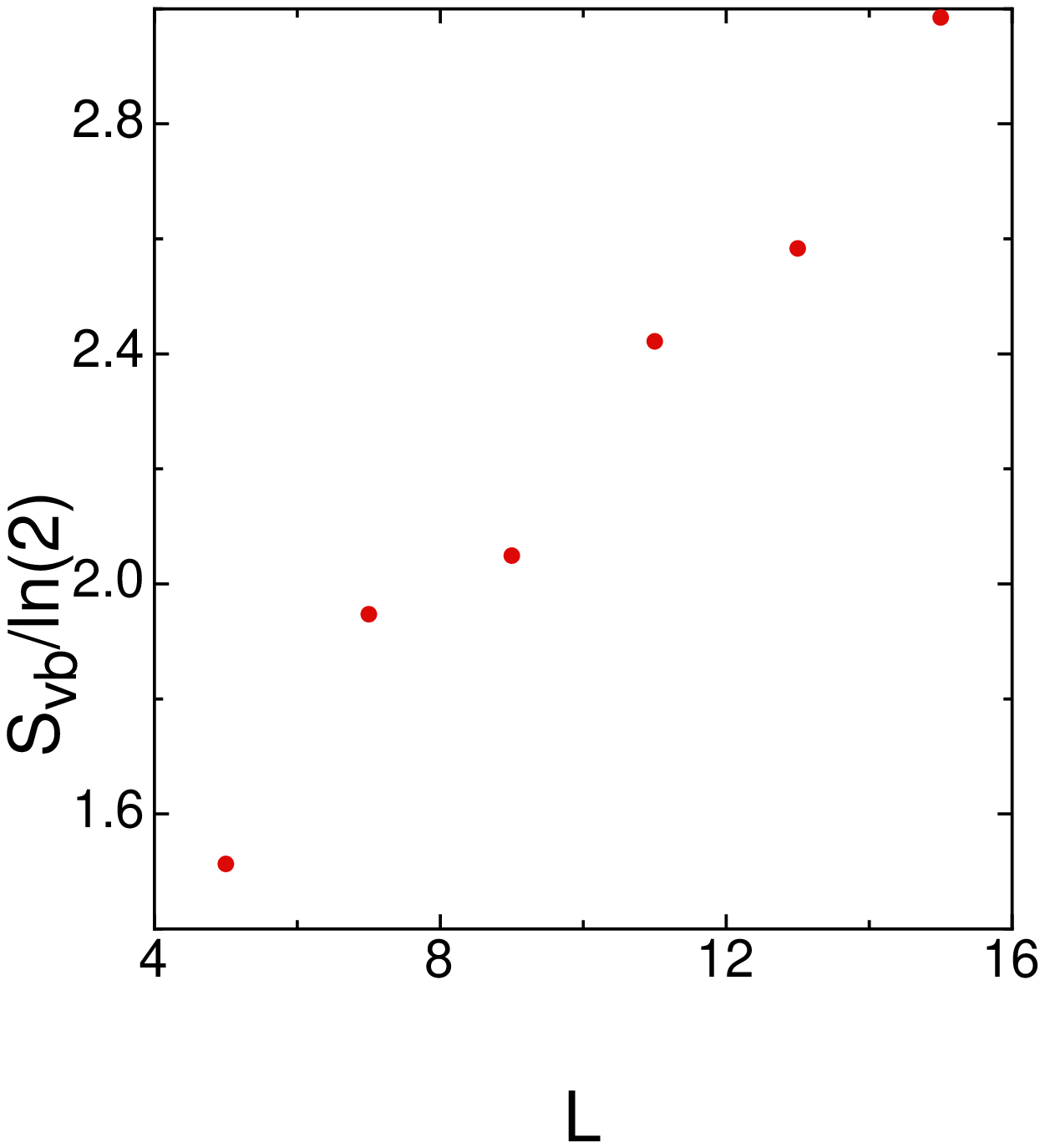}
\caption{Valence Bond Entropy vs. system size for a bisected system, $L$ odd.}
\label{fig.11a}
\end{figure}

Figure 10 is a plot of the second Renyi entropy $S_2$ vs. $L$ for $L$ even taking the subsystem as half the cluster calculated using valence bond Monte Carlo.  The values of $S_2$ are noticeably smaller than for comparably sized Bethe Clusters.  However, $S_2$ does not appear to be a constant or depend on $\ln L$.  The calculation appears, to within the large statistical errors and the limited system sizes, to be consistent again with $S_2$  scaling as $L$.  To make the case for $L$ more convincing, figure 11 is a plot of valence bond entropy vs. $L$ for various system sizes and $L$ even.  One sees a rather clear indication of the valence bond entropy scaling with $L$.  The same quantities are plotted for $L$ odd in figure 12.  Again one sees the valence bond entropy scaling as $L$.  The structure in the graph can be rationalized by noting that for $L = 2n+1$ $n$ even, there is a valence bond pattern with only one valence bond joining the two $ L \times L$ squares and that bond can be chosen to be the middle bond  (where the interaction joining the squares is located).

\section{Conclusions}
Numerical evidence has been presented that for a Heisenberg model on clusters joined by a single interaction, the entanglement entropy scales as the perimeter, not as a constant, as suggested by the area law.  This is consistent with work on non interacting fermions and demonstrates these earlier results are not an artifact of non interacting particles.  As discussed in the introduction, this has ramifications for DMRG calculations on Bethe clusters.  More generally, the scaling with perimeter places fundamental limitations on DMRG algorithms involving partitioning into clusters.  This is analogous to the restrictions placed on heat engines by the thermodynamic entropy  where "heat engine"  in this case is a quantum calculation (done on a classical computer) and "entropy" is the entanglement entropy.   In addition, two $L \times L$ clusters joined by a link is an experimentally realizable system.  To directly measure entanglement entropy is difficult (though not impossible \cite{islam} ).  However, recent work on the area law suggests the obstruction to the area law in two dimensions ( see \cite{marien} p34-39 ) is the existence of edge states.   Such (hypothetical) edge states may be easier to measure (and analyze) than directly measuring the entanglement entropy.  As well as telling us what we cannot do, insights from the second law of thermodynamics can be used for more rational engine design.  An analogous situation could conceivably hold for the "area" law, where now the device analogous to a heat engine is a quantum computer.  Proposed modular quantum computer architectures\cite{monroe1,monroe2} are quite similar to the clusters joined by links considered in the present work.

\end{document}